\documentclass[10pt]{article}

\usepackage{amsmath}
\usepackage{amssymb}

\usepackage{graphicx}

\usepackage{cite}

\usepackage{color} 

\usepackage[labelfont=bf,labelsep=period,justification=raggedright]{caption}

\makeatletter
\renewcommand{\@biblabel}[1]{\quad#1.}
\makeatother

\date{}

\definecolor{gray}{rgb}{0.6,0.6,0.6}

\begin{document}


\begin{flushleft}
{\Large
\textbf{How to understand the cell by breaking it: network analysis of gene perturbation screens}
}
\\
Florian Markowetz$^\ast$
\\
Cancer Research UK Cambridge Research Institute, Cambridge CB2 0RE, UK\\
$\ast$ florian.markowetz@cancer.org.uk
\end{flushleft}

\section*{Abstract}

Modern high-throughput gene perturbation screens are key technologies at the forefront of
genetic research. Combined  with rich phenotypic descriptors they enable researchers to observe detailed cellular reactions to experimental perturbations on a genome-wide scale.
This review surveys the current state-of-the-art in analyzing single gene perturbation screens from a network point of view. 
We describe approaches to make the step from the parts list to the wiring diagram by using phenotypes for network inference and integrating them with complementary data sources. 
The first part of the review describes methods to analyze one- or low-dimensional phenotypes like viability or reporter activity; the second part concentrates on high-dimensional phenotypes showing global changes in cell morphology, transcriptome or proteome.


\section*{Introduction}

Functional genomics has demonstrated considerable success in inferring the inner working of a cell through analysis of its response to various perturbations. In recent years several technological advances have pushed gene perturbation screens to the forefront of functional genomics.  Most importantly, modern technologies make it possible to probe gene function on a genome-wide scale in many model organisms and human. For example, large collections of knock-out mutants  play a prominent role in the study of \emph{S. cerevisiae} \cite{Winzeler1999} and RNA interference (RNAi) has become a widely used high-throughput method to knock-down target genes in a wide range of organisms, including \emph{Drosophila melanogaster}, \emph{C. elegans}, and human \cite{Fuchs2006,Moffat2006,Boutros2008}. 

Another major advance is the development of rich phenotypic descriptions by imaging or measuring molecular features globally. Observed phenotypes can reveal which genes are essential for an organism, or work in a particular pathway, or have a specific cellular function.  Combining high-throughput screening techniques with rich phenotypes enables researchers to observe detailed reactions to experimental perturbations on a genome-wide scale. This makes gene perturbation screens one of the most promising tools in functional genomics.

Advances in the design and analysis of gene perturbation screens may have an immediate impact on many areas of biological and medical research. New screening and phenotyping techniques often directly translate into new insights in gene and protein functions. Results of perturbation screens can also reveal unexploited areas of potential therapeutic intervention. For example, a recent RNAi screen showed that some of the most critical protein kinases for the proliferation and survival of cancer cell lines are also the least studied \cite{Luo2008}. 

A goal becoming more and more prominent in both experimental as well as computational research is to leverage gene perturbation screens to the identification of molecular interactions, cellular pathways and regulatory mechanisms. Research focus is shifting from understanding the phenotypes of single proteins to understanding \emph{how} proteins fulfill their function, \emph{what} other proteins they interact with and \emph{where} they act in a pathway.  Novel ideas on how to use perturbation screens to uncover cellular wiring diagrams can lead to a better understanding of how cellular networks are de-regulated in diseases like cancer. This knowledge is indispensable for finding new drug targets to attack the drivers of a disease and not only the symptoms.

\paragraph{Phenotypes} 
A phenotype can be any observable characteristic of an organism. Analysis strategies strongly depend on how rich and informative phenotype descriptors are. We will call phenotypes resulting from a single reporter (or a small number of reporters) \emph{low-dimensional} phenotypes and the genes showing significant results \emph{hits} \cite{Boutros2006,Rieber2009}. Examples of such low-dimensional phenotypes are cell viability versus cell death \cite{Winzeler1999}, growth rates \cite{Giaever2002} or the activity of reporter constructs, e.g.~a luciferase, downstream of a pathway of interest \cite{Mueller2005}. Low-dimensional phenotyping screens can identify candidate genes on a genome-wide scale and are often used as a first step for follow-up analysis. We will discuss methods to functionally interpret hits from low-dimensional phenotyping screens and to place them in the context of cellular networks in the first part of this review. 

The second part will be devoted to \emph{high-dimensional} phenotyping screens, which evaluate a large number of cellular features at the same time. Observing system-wide changes promises key insights into cellular mechanisms and pathways that can not be supplied by low-dimensional screens. For example, high-dimensional phenotypes can include changes in cell morphology  \cite{Perlman2004,Gunsalus2004,Neumann2006,Bakal2007}, or growth rates under a wide range of conditions \cite{Brown2006}, or transcriptional changes measured on microarrays \cite{Hughes2000,Boutros2002,Ivanova2006,Amit2009}, or changes in the metabolome and proteome \cite{Ideker2001} measured by mass spectrometry \cite{Gstaiger2009} or flow cytometry \cite{Sachs2005,Niu2008}. Morphological and growth phenotypes can be obtained on a genome-wide scale \cite{Bakal2007,Brown2006}, while transcriptional and proteomic phenotypes  are often restricted to individual pathways or  processes \cite{Boutros2002,Sachs2005,Ivanova2006}.

The distinction between low- and high-dimensional phenotypes may sound technical, but it is crucial for choosing potential analysis methods. The central difference is that high-dimensional phenotypes allow to compute correlations and other similarity measures, which are not applicable for low-dimensional phenotypes.  Another important distinction is between \emph{static} phenotypes, providing a `snapshot' of a cell's reaction to a gene perturbation, and \emph{dynamic} phenotypes showing a cell's reaction over time.  We expect more and more studies in the future to produce dynamic output and in the following note explicitly which methods can be applied to dynamic phenotypes. For the biological interpretation of screening results it is very important to keep in mind which level of `cellular granularity' a phenotype describes: growth rates or cell morphologies are much more `high-level' features of the cell than gene or protein expressions. As soon as more studies produce dynamic phenotypes on many different cellular levels, integrative analysis of inter-connected phenotypes \cite{Lu2009}  will become more important.  In the following, however, we concentrate on the current state-of-the art, which almost always uses a single type of readout in a perturbation screen.

\paragraph{Pre-processing pipeline}
In this review we focus on single gene perturbations by knockouts \cite{Winzeler1999} or RNA interference \cite{Boutros2008} that allow targeting individual genes or combinations of genes. 
Before network analysis, the raw data needs to pass an initial analysis and quality control pipeline specific to the perturbation and phenotyping technologies used.
Low-dimensional screens are mostly performed in multiple-well-plates  and a typical analysis pipeline \cite{Boutros2008} includes data pre-processing, removal of spatial biases per plates, normalization between plates, and finally detection of significant hits \cite{Boutros2006,Rieber2009,Birmingham2009}.  In vertebrates, genes need to be targeted with multiple siRNAs to ensure effective down-regulation \cite{Boutros2008} and the multiple phenotypes per gene can afterwards be integrated into a statistical score \cite{Konig2007}. High-dimensional morphological screens depend on computational analysis like image segmentation \cite{Carpenter2006,EBImage} and phenotype discovery \cite{Jones2009,Yin2008,Wang2008} for rapid and consistent phenotyping. Molecular high-dimensional phenotypes need pre-processing depending on their platform and different approaches exist e.g. for flow-cytometry data \cite{Hahne2006} or microarrays  \cite{Smyth2005}. 

\paragraph{From phenotypes to cellular networks} 
The phenotypes we have discussed above allow only an  indirect view on how different genes in the same process interact to achieve a particular phenotype. Cell morphology or sensitivity to stresses, for example, are global features of the cell and hard to relate directly to how individual genes contribute to them (see Fig.~1a). Gene expression phenotypes show transcriptional changes in the genes downstream of a perturbed pathway    but offer only an indirect view of pathway structure due to the high number of non-transcriptional regulatory events like protein modifications \cite{Markowetz2005b}. For example,  different protein activation states by phosphorylation may not be visible by changes in mRNA concentrations (see Fig.~1b). 

This gap between observed phenotypes and underlying  cellular networks is the main problem in the analysis of perturbation screens and applies to both low- and high-dimensional screens. 
The goal of computational analysis is to bridge this gap by inferring gene function and recovering pathways and mechanism from observed phenotypes. The following methods address the challenge in different ways, mostly by integrating the perturbation effects and phenotypes with additional sources of information like collections of functionally related gene sets or protein-interaction networks.

\section*{Network analysis of low-dimensional phenotypes}

\paragraph{Global overview by enrichment analysis} 
A simple way to link phenotypes to gene function is to test whether pathways or functional groups of genes  (e.g. defined by Gene Ontology terms \cite{Ashburner2000} or  MSigDB \cite{Subramanian2005}) are enriched in the list of hits. Most methods use a hypergeometric test statistic (see Fig.~2a) and many can be used online \cite{Huang2009,Sealfon2006,Bauer2008} or as Bioconductor packages \cite{Alexa2006}. An alternative global functional annotation method tests whether functional groups show a trend towards especially strong or weak phenotypes without using a cutoff to define hits \cite{Subramanian2005} (see Fig.~2b). Enrichment analysis can also be very useful  to analyze high-dimensional phenotypes, for example when functionally annotating the results of a clustering method.

Enrichment analysis results in a list of $p$-values  describing how significantly each gene set was represented in the hits. Enrichment analysis reduces complexity and improves interpretability of results by moving from single genes to functionally related gene sets. This type of analysis is often called 'un-biased' and 'hypothesis-free' and is ideal for a comprehensive first overview. 
However, enrichment analysis loses its value for complexity reduction if the number of gene sets becomes too big. Also, overlap and dependencies between gene lists that could potentially bias the results have so far only been addressed for the GO graph \cite{Alexa2006,Bauer2008} but not for more general collections of gene lists like MSigDB \cite{Subramanian2005}.

Good data analysis asks specific questions.  A Ôhypothesis-freeÕ method can only be the very first starting point for a deeper exploration of the data. For example, all enrichment methods rely on known gene sets and cannot uncover new pathways or components. Enrichment methods treat pathways as bags of unconnected genes without considering connections within and between pathways. Thus, enrichment methods can only deliver a very crude picture of the cell. In the following we will discuss approaches to overcome some of the limitations of enrichment analysis by integrating the observed phenotypes with complementary sources of information.

\paragraph{Mapping phenotypes to networks}
Another valuable source of information  to interpret RNAi hits are gene and protein networks obtained either experimentally \cite{Bork2004,Stelzl2005} or computationally by literature mining \cite{Ma'ayan2005} or integrating heterogeneous genomic data \cite{Lee2004,Myers2005,Guan2008}. All computational networks are available online on supplementary webpages and the experimental networks can be obtained from databases like STRING \cite{Jensen2009} or BioGRID \cite{Breitkreutz2008}.

Using these complementary data sources can improve hit identification \cite{Kaplow2009,Wang2009,Berndt2009} and even provide a more refined view of the pathways the hits contribute to. One strategy is  to search for sub-networks containing a surprisingly large number of hits (see Fig.~3a). While this strategy is already useful when evaluating interesting sub-networks by eye \cite{Lee2008,Krishnan2008} its true power comes from the availability of efficient  search algorithms to find sub-networks enriched for RNAi hits  and assess their significance \cite{Ideker2002,Konig2008,Dittrich2008,Bankhead2009,Tu2009}. An additional application of mapping hits to a network is that known phenotypes can be used to predict phenotypes of genes not included in the screen, e.g. by assuming that a gene connected to many hits should also show a strong phenotype \cite{Lee2008}. The success of all network-mapping strategies strongly depends on the quality and coverage of both the screen and the linkage in the network.

\paragraph{Gene prioritization} 
Other approaches complement genomic data with biological prior knowledge showing how `interesting' hits look like. Gene prioritization \cite{Wang2009,Aerts2006}  ranks genes according to how promising they would be for follow-up studies. Because it uses prior knowledge to fine-tune the  algorithm, gene prioritization can be more focussed than a global un-informed search for enriched subnetworks.

\section*{Network analysis of high-dimensional phenotypes}

\paragraph{Global overview by clustering and ranking.} 
Most state-of-the-art analysis techniques rely on a Ôguilt-by-associationÕ paradigm: genes with similar phenotypes will most probably have a similar biological function. This explains the prevalence of clustering techniques  in analyzing high-dimensional phenotyping screens \cite{Perlman2004,Brown2006,Ivanova2006,Bakal2007}. Clustering is a convenient first analysis and visualization step that can can highlight strong trends and patterns in the data and can thus yield a global first impression of functional units. Another analysis strategy relying on guilt-by-association is to rank genes by their phenotypic similarity compared to a gene of interest \cite{Gunsalus2004}. Clustering and ranking can be combined with enrichment analysis (as discussed above)  for functional interpretation.

\paragraph{Graph methods linking causes to effects}  
Another useful data visualization especially for transcriptional phenotypes is to build a directed (not necessarily acyclic) graph by drawing an arrow between two genes if perturbing one results in a significant expression change at the other \cite{Rung2002}. This graph representation  can be then used as a starting point for further analysis, for example by using graph-theoretic methods of transitive reduction \cite{Aho1972} to  distinguish between direct and indirect effects of a perturbation \cite{Wagner2001,Tresch2007}.

\paragraph{Probabilistic graphical models.} 
Most approaches to infer pathway structure from experimental data rely on probabilistic graphical models. For low-dimensional phenotypes they often suffer from non-uniqueness and un-identifiability issues  \cite{Kaderali2009}, but can be applied very successfully in high-dimensional settings.  A prominent approach are (static or dynamic) Bayesian networks, which describe probabilistically how a gene is controlled by its regulators \cite{Friedman2004,Markowetz2007a}.  To model experimental perturbations most approaches rely on the  concept of `ideal interventions' \cite{Pearl2000} which deterministically fix a target gene to a particular state (e.g `0' for a gene knockout). Ideal interventions were applied in Bayesian networks \cite{Ellis2008,Sachs2005,Peer2001}, factor graphs \cite{Gat-Viks2006} and dependency networks \cite{Rogers2005}. In simulations \cite{Werhli2006,Zak2003} and on real data \cite{Sachs2005,Werhli2006} it was found that interventions are critical for effective inference.

The model of ideal interventions contains a number of idealizations (hence the name), most importantly that manipulations only affect single genes and that perturbation strength can be controlled deterministically. The first assumption may not be true if there are off-target  or compensatory effects involving other genes. The second assumption may also not hold true in realistic biological scenarios; in particular for RNAi screens where experimentalists often lack knowledge about the exact knock-down efficiency. Probabilistic generalizations of ideal interventions can be used to cope with this uncertainty \cite{Markowetz2005a}. 

\paragraph{Probabilistic data integration}
High-dimensional phenotypic profiles can be mapped to given graphs and networks by finding subgraphs that are connected in the background network and at the same time show high similarity of phenotypic profiles. These approaches already exist for mapping gene expression data onto protein interaction networks \cite{Ulitsky2007} and the same algorithms could easily be applied to any other kind of high-dimensional phenotypic profiles (see Fig.~3b). Other approaches use data integration to construct potential pathways from protein interactions and transcription factor binding data to relate perturbed genes to the observed downstream effects \cite{Yeang2004,Yeang2005,Ourfali2007}.

\paragraph{Multiple Input - Multiple Output (MIMO) models}
Many of the approaches discusses so far--like clustering or graphical models--can be applied to both static `snapshots' as well as dynamic time-course measurements. 
Another approach  to model specifically the dynamics of networks comes from a branch of control theory called `systems identification'  \cite{Ljung1986} and uses so called Multiple Input - Multiple Output (MIMO) models.
MIMO models  represent the evolution of a perturbed cell over time by linear differential equations \cite{Tegner2003,Gardner2003,Xiong2004,diBernardo2005,Lorenz2009}  and can represent non-linear effects by transfer functions \cite{Nelander2008}. The models can be inferred by regression techniques in the linear case \cite{Gardner2003} or Monte Carlo stochastic search in the non-linear case \cite{Nelander2008}. The framework is very flexible and can incorporated single as well as combinatorial perturbations.

\paragraph{Nested Effects Models (NEMs)} 
One of the key problems in analyzing perturbation screens is that the observed phenotypes are downstream of the perturbed pathway and may not show the direct influence of one pathway component on another. A class of models explicitly addressing this problem are Nested Effects Models \cite{Markowetz2005b,Markowetz2007}. They reconstruct pathway structure from \emph{subset relations} based on the following rationale: Perturbing some genes may have an influence on a global process, while perturbing others affects sub-processes of it. Imagine, for example, a signaling pathway activating several transcription factors. Blocking the entire pathway will most probably affect all targets of all transcription factors, while perturbing a single  transcription factor will only affect its direct targets, which are a subset of the phenotype obtained by blocking the complete pathway. Given high-dimensional phenotypes showing a subset structure, NEMs find the most likely pathway topology explaining the data. They differ from other statistical approaches like Bayesian networks by encoding subset relations instead of correlations or other similarity measures. The theory of NEMs has been applied and extended in several studies \cite{Frohlich2007,Tresch2008,Frohlich2008b,Frohlich2009}. An implementation is available as an R/Bioconductor package \cite{Frohlich2008a}. Other extensions to the NEM framework distinguish between activating and inhibiting regulation \cite{Vaske2009}  or include dynamic information from time-series measurements \cite{Anchang2009}.

\section*{Discussion and Outlook}

In this review we have discussed two main approaches to describe the reaction of a cell to an experimental gene perturbation: low-dimensional phenotypes measure individual reporters for cell viability or pathway activation, while high-dimensional phenotypes show global effects on cell morphology, transcriptome or proteome.  Table~1 lists examples of freely available software implementing some of these approaches. All of them can be directly applied to gene perturbation screens, even though some of them have been introduced in different contexts. While this review has focused on single gene knock-outs and knock-downs, similar approaches can  be applied to gene over-expression screens  \cite{Sopko2006,Stokic2009,Niu2008,Lorenz2009}, drug treatment \cite{Nelander2008}, environmental stresses changing many genes \cite{Yosef2006,MacCarthy2005} or even natural genetic variation \cite{Rockman2008}. 

\paragraph{Predicting phenotypes from metabolic networks}
The focus of this review is on functionally annotating hits in a network context and reconstructing networks from high-dimensional phenotypes. In a complementary direction of research, genome-wide reconstructions of metabolic networks \cite{Herrgaard2008,Duarte2007} are used to predict effects of gene perturbations. Instead of predicting networks from phenotypes, these approaches predict phenotypes from networks. 
For example, in \emph{S. cerevisiae} and \emph{E. coli}  computational models very accurately predict fitness effects of gene knock-outs  \cite{Papp2004,Fong2004} as well as  compensatory rescue effects \cite{Motter2008}. 
However, recent developments in metabolic network modeling have led to linear programming algorithms to extract relevant context-specific sub-networks of activity from a genome-wide network \cite{Shlomi2008,Becker2008}. In the same way as the probabilistic data integration methods discussed above, e.g. \cite{Ulitsky2007},  these algorithms could be used in the future to find metabolic sub-networks active under certain gene perturbations. 

\paragraph{From single to combinatorial perturbations} 
While single gene perturbation screens have been immensely successful in extending our knowledge of pathway components and interactions,  an important limitation can be caused by compensatory effects, genetic buffering and redundancy of cellular mechanisms and pathways \cite{Deutscher2008,Gitter2009}. 
This can only be overcome by perturbing several genes at the same time. The number of possible combinations grows rapidly and thus current approaches are mainly limited to perturbing pairs of genes and observing low-dimensional phenotypes like fitness estimates \cite{Tong2004}.  The analysis of combinatorial perturbations is the topic of another review \cite{Myers2009}.

\paragraph{The end of the screen is the beginning of the experiment}  Global phenotyping and pathway screening can be combined in the same study. For example, a first genome-wide screen identifies key genes representative for pathways and cellular mechanisms involved in the phenotype. In a second step the hits of the first screen could be assayed for high-dimensional molecular phenotypes to infer a pathway diagram using Nested Effects Models or other statistical approaches. 

In a further step this preliminary pathway models could be used to plan an additional round of experimentation. Different modeling frameworks propose future experiments to most effectively refine a pathway hypothesis, e.g. Bayesian networks \cite{Pournara2004,Yoo2004}, physical network models \cite{Yeang2005}, logical models \cite{Szczurek2009}, Boolean networks \cite{Ideker2000}, and dynamical modeling \cite{Tegner2003}.

Iteratively integrating experimentation and computation may lead to a virtuous circle and is one of the most promising approaches to refine our understanding of the inner working of the cell.

\section*{Acknowledgments}

I thank the organizers of the ISMB 2009 tutorial sessions for the opportunity to present this review. Yinyin Yuan, Roland Schwarz, Gregoire Pau provided helpful comments on drafts of the manuscript. My research is funded by Cancer Research UK.


\newpage
\section*{Figures and  legends}

\begin{figure}[!ht]
\begin{center}
\includegraphics[width=\textwidth]{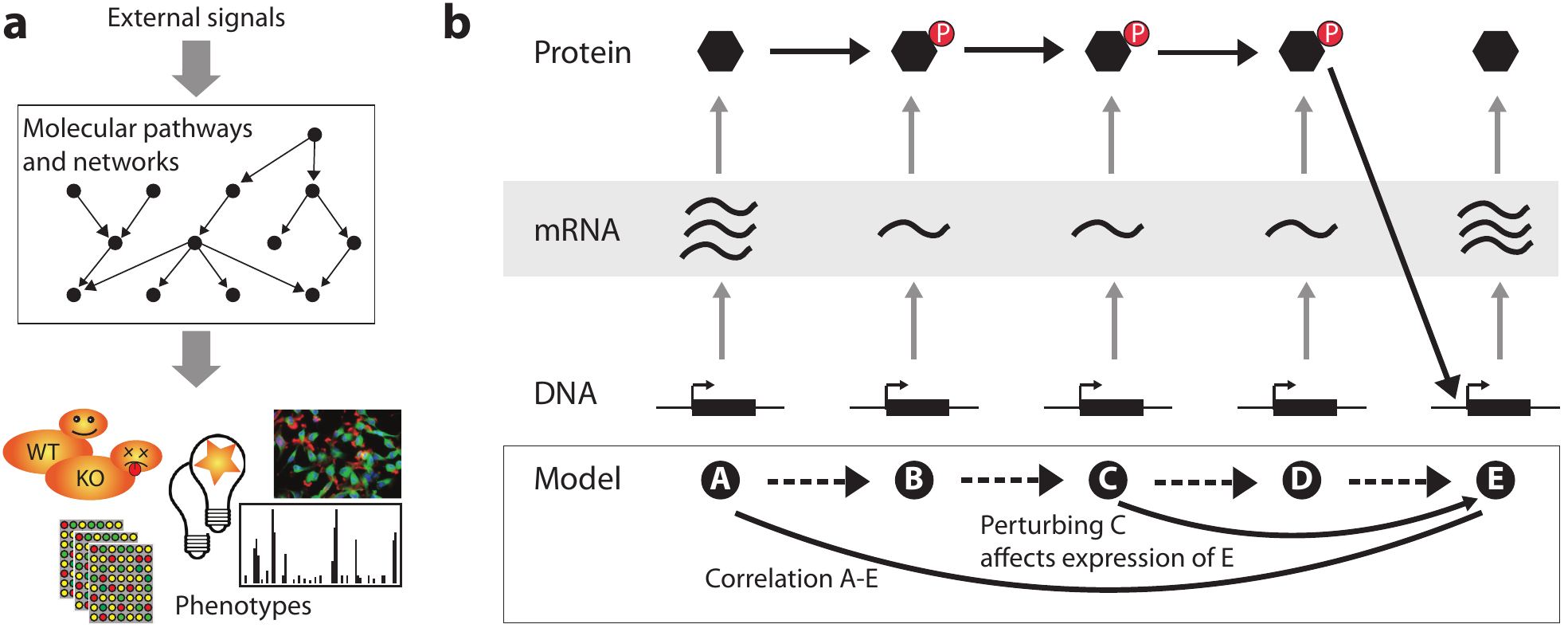}
\end{center}
\caption{
{\bf Cellular networks underlying observable phenotypes.}   \textbf{(a)} Phenotypes are the response of the cell to external signals mediated by cellular networks and pathways. The goal of computation is to reconstruct these networks from the observed phenotypes.  \textbf{(b)} Global molecular phenotypes like gene expression allow a view inside the cell but also have limitations. This is exemplified here in a cartoon pathway adapted from \cite{Wagner2001} showing a cascade of five genes/proteins (A-E). Proteins A-C form a kinase cascade, D is a transcription factor acting on E. Up-regulation of A starts information flow in the cascade and results in E being turned on. In gene expression data this is visible as a correlation between A and E (represented as an undirected edge in the model). Experimentally perturbing a genes, say C, removes the corresponding protein from the cascade, breaks the information flow and results in an expression change at E (represented as an arrow in the model). However, the different phosphorylation and activation states of proteins B-D will most probably not be visible as changes in gene expression. Thus, due to the pathway mostly acting on the protein level most parts of the cascade (dashed arrows in the model) can not be inferred from gene expression data directly.}
\label{Figure1}
\end{figure}

\vfill

\begin{figure}[!ht]
\begin{center}
\includegraphics[width=\textwidth]{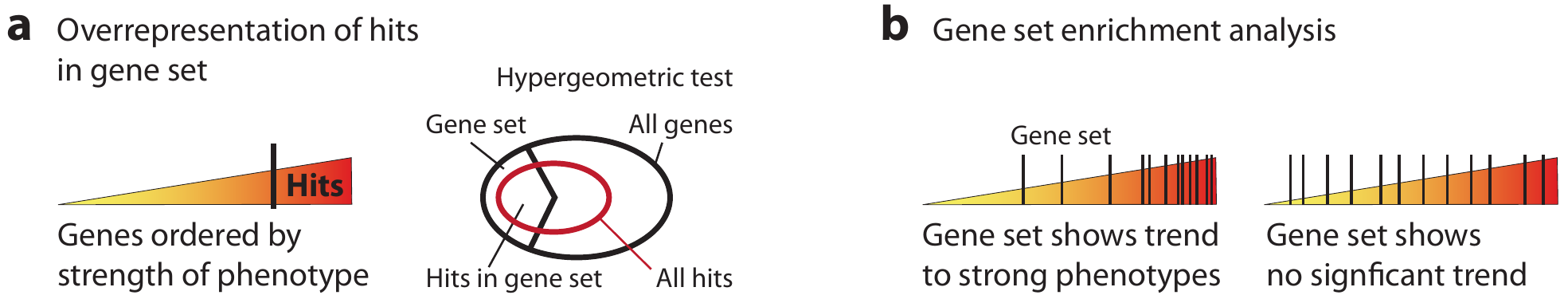}
\end{center}
\caption{
{\bf Functional annotation of hits by enrichment analysis.}  \textbf{(a)} In the first approach  \cite{Bauer2008} a cutoff is applied to select the hits with strongest phenotypes. A hyper-geometric test then evaluates if the overlap between the hits and a given gene set is surprisingly large (or small) compared to the overlap with a random set. \textbf{(b)} A second approach \cite{Subramanian2005} does not need a cutoff. It maps the gene set (black bars) onto the observed phenotypes and quantifies if there is a significant trend or if the genes are spread out uniformly over the whole range.}
\label{Figure2}
\end{figure}

\begin{figure}[!ht]
\begin{center}
\includegraphics[width=\textwidth]{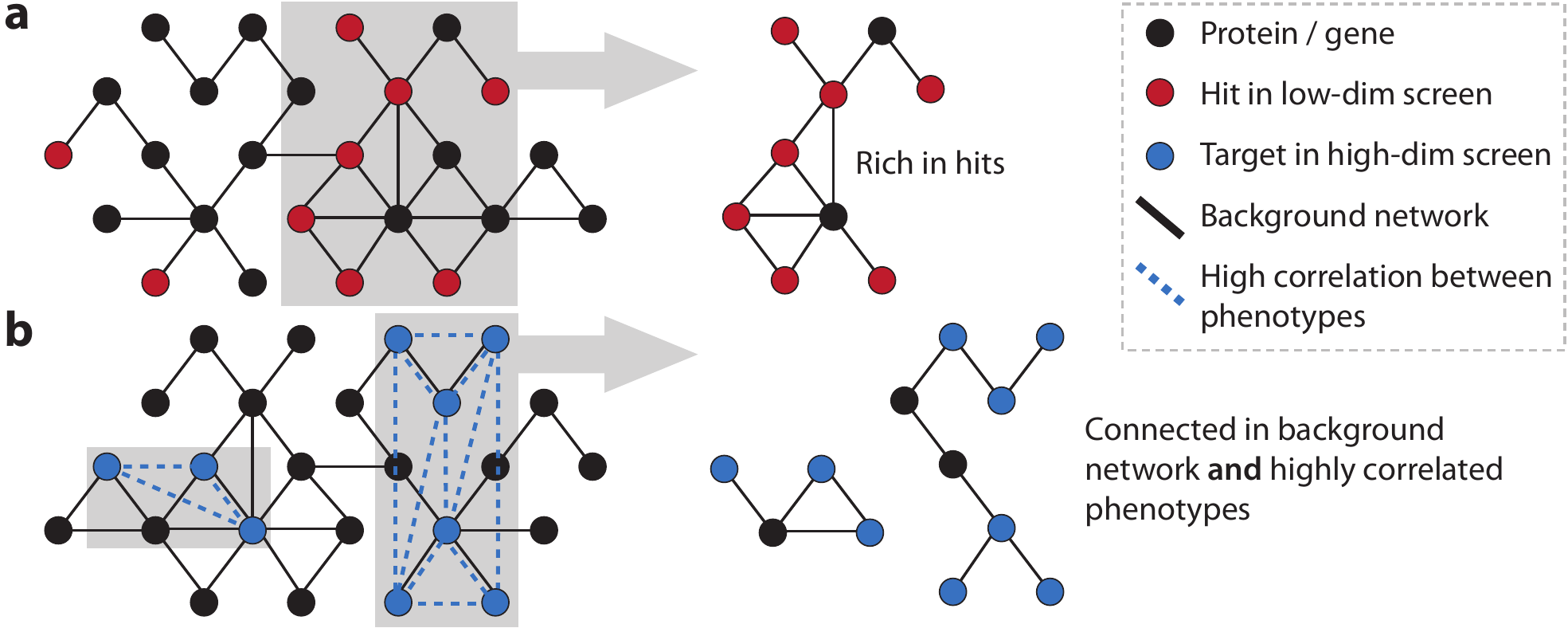}
\end{center}
\caption{
{\bf Extracting rich sub-networks.}  Different patterns in the graph point to a common cellular mechanism causing a phenotype: \textbf{(a)} hits in a low-dimensional screen (red nodes) clustering in highly connected sub-networks, and  \textbf{(b)}  high correlation between high-dimensional phenotypes of target genes connected in the background network. The black graph represents any type of background network. }
\label{Figure3}
\end{figure}

\newpage

\section*{Tables}

{\small
\begin{table}[!ht]
\caption{\bf Examples of software for network analysis of gene perturbation screens. }
\begin{tabular}{|p{2cm}|p{8cm}|l|}
\hline
\multicolumn{3}{l}{\bf General data analysis and network visualization}\\
\hline
Bioconductor & Software environment for the analysis of genomic data featuring hundreds of contributed packages  \cite{Bioconductor}  &  www.bioconductor.org \\
Cytoscape &   Software platform for visualizing molecular interaction networks and integrating them with other data types \cite{Shannon2003} &  www.cytoscape.org \\
\hline\hline
\multicolumn{3}{l}{\bf Setting up data for network analysis}\\
\hline
cellHTS2 &   End-to-end analysis of cell-based screens: from raw intensity readings to the annotated hit list \cite{Boutros2006} & www.bioconductor.org \\
RNAither &   Analysis of cell-based RNAi screens,  includes quality assessment and customizable normalization \cite{Rieber2009} & www.bioconductor.org\\
EBImage &   Cell image analysis and feature extraction \cite{EBImage} & www.bioconductor.org \\
CellProfiler &   Cell image analysis and feature extraction \cite{Carpenter2006}& www.cellprofiler.org\\
\hline\hline
\multicolumn{3}{l}{\bf Enrichment analysis}\\
\hline
DAVID &  Tools for data annotation, visualization and integration \cite{Huang2009} & david.abcc.ncifcrf.gov \\
GOLEM &   Enrichment analysis and visualization of GO graph (Fig 2a) \cite{Sealfon2006} &  function.princeton.edu/GOLEM \\
Ontologizer &  Enrichment analysis with dependencies between GO nodes (Fig 2a) \cite{Bauer2008} &   compbio.charite.de/ontologizer\\
GSEA &  Gene set enrichment analysis (Fig 2b) \cite{Subramanian2005}& www.broadinstitute.org/gsea/\\
\hline\hline
\multicolumn{3}{l}{\bf Clustering and ranking}\\
\hline
Cell Profiler Analyst   & Interactive exploration and analysis of multidimensional data from image-based experiments \cite{Jones2009} &  www.cellprofiler.org \\
PhenoBlast  & Ranking of phenotype profiles according to similarity with given profile \cite{Gunsalus2004} & www.rnai.org \\
Endeavour & Prioritizes hits for further analysis \cite{Aerts2006} & www.esat.kuleuven.be/endeavour/ \\ 
\hline\hline
\multicolumn{3}{l}{\bf Finding rich sub-networks}\\
\hline
heinz  & Finds optimal subnetworks rich in hits (Fig 3a)\cite{Dittrich2008}  & www.planet-lisa.net\\
jActiveModules & Finds heuristic subnetworks rich in hits (Fig 3a) \cite{Ideker2002} & www.cytoscape.org \\
Matisse  & Finds subnetworks with high phenotypic similarity (Fig 3b) \cite{Ulitsky2007} & acgt.cs.tau.ac.il/matisse/ \\
\hline\hline
\multicolumn{3}{l}{\bf Network reconstruction}\\
\hline
nem  & Nested Effects Models reconstruct pathway features from subset relations in high-dim phenotypes  \cite{Frohlich2008a} & www.bioconductor.org \\
copia & Copia uses MIMO models to reconstruct networks from perturbations \cite{Nelander2008} & cbio.mskcc.org/copia/  \\
\hline
\end{tabular}
\begin{flushleft}
The table contains the name of the method, a short description with reference, and a webpage where it can be obtained. This list is far from comprehensive, but hopefully provides a starting point even for non-coding experimentalists.
\end{flushleft}
\label{tab:label}
\end{table}
}


\end{document}